\documentclass{article}
\usepackage{xspace,amssymb,amsmath,helvet,graphicx,url}
\makeatletter
\renewcommand\@biblabel[1]{}
\makeatother

\newcommand{\dee}{\,\mbox{d}}

\newcommand{\eg}{e.g.\xspace}
\newcommand{\ie}{i.e.\xspace}

\newcommand{\etc}{etc.\@\xspace}

{\begin{list}{}{%
\settowidth{\labelwidth}{\textsf{#1:}}%
\setlength{\leftmargin}{\labelwidth}
\addtolength{\leftmargin}{\labelsep}
\setlength{\itemindent}{0pt}
}}%
{\end{list}}
\begin{document}
\title{Properties and Applications of some Distributions derived from Frullani's integral\\
Rose Baker\\Centre for Operational Research and Applied Statistics\\University of Salford, UK\\r.d.baker@salford.ac.uk}
\maketitle
\begin{abstract}
Frullani's integral dates from 1821, but a probabilistic
interpretation of it has never been made. In this paper,
Frullani's integral formula is shown to result from mixing a
lifetime distribution by allowing the logarithm of the scale
factor to be uniformly distributed over a finite range. This
gives a class of long-tailed distributions related to slash
distributions, where the pdf is simply expressed in terms of the
survival function of the `parent' distribution.  The resulting
survival distributions have all moments finite, and can exhibit
the bimodal hazard functions sometimes seen in practice. A
distribution of this type analogous to the t-distribution is
derived, the corresponding multivariate distributions are
given, and two skewed versions of this distribution are derived.
The use of the mixed distributions for inference is exemplified
by fitting them to several datasets. It is expected that there
will be many applications, in health, reliability,
telecommunications, finance, etc.
\end{abstract}
\section*{Keywords}
Bimodal hazard, Frullani integral, hazard function, mixture distribution,
survival distribution, two-piece distribution.
\section{Introduction}
Frullani's result is that
\begin{equation}\int_0^\infty \frac{F(at)-F(bt)}{t}\dee t=\ln(a/b)\{F(\infty)-F(0)\},\label{eq:frullani}\end{equation}
where $F(0)=\lim_{t\rightarrow
0}F(t)$, $F(\infty)=\lim_{t\rightarrow \infty}F(t)$, $ a > 0, b
> 0$ and where
$F(t)$ is Lebesgue-integrable in $(0,\infty)$. Arias-De-Renya (1990) gives
the fullest account of the history of the discovery of this result,
which was first given by Frullani in 1821, and later by Cauchy in
1823 and 1827. Iyengar (1941) gives the
first modern analysis, followed by Ostrowski (1949), Agnew (1951),
Ostrowski (1976) and Arias-De-Renya (1990). The modern work is largely
concerned with replacing the limits $F(\infty)$ and $F(0)$ by
suitable mean values, and is not directly relevant to this paper.

One's initial reaction is probably surprise that the integral is
determined by $F(0)$ and $F(\infty)$ alone. Both terms in
the integrand are infinite, and a quick proof of
(\ref{eq:frullani}) rigorous enough for most proceeds by
removing this problem by evaluating
$\int_{\epsilon}^{1/\epsilon}\{F(at)/t\}\dee
t-\int_{\epsilon}^{1/\epsilon}\{F(bt)/t\}\dee t$ for small
$\epsilon$, and then letting $\epsilon\rightarrow 0$. This is
left for the reader's amusement, as is another loose proof based
on integration by parts.

However, the connection with survival distributions follows from
a different proof of (\ref{eq:frullani}), which is rewritten as
\begin{equation}\frac{1}{\ln(a/b)}\int_0^\infty \frac{F(at)-F(bt)}{t}\dee t=1,\label{eq:probint}\end{equation}
where, without loss of generality,  $a > b$, and where $F$ is the
distribution function of a survival distribution. The meaning is
now that 
\begin{equation}g(t)=\frac{F(at)-F(bt)}{\ln(a/b)t}\label{eq:pdf}\end{equation}
is a pdf where $T > 0$ if $F$ is a monotonically increasing function of
its argument; naturally,
$F$ must obey this condition to be a valid distribution
function. The pdf $g$ then
integrates to unity and is everywhere non-negative.  As $b
\rightarrow a$, $g(t)$ tends to the pdf $f(t|a)$ of $F(at)$.
We can also write (\ref{eq:pdf}) in terms of the survival
functions $\bar{F}=1-F$, when
\begin{equation}g(t)=\frac{\bar{F}(bt)-\bar{F}(at)}{\ln(a/b)t}.\label{eq:spdf}\end{equation}
To explore the meaning of this pdf, we write
\begin{equation}\frac{F(at)-F(bt)}{\ln(a/b)t}=\frac{\int_b^a (\dee F(ut)/\dee u)\dee u}{\ln(a/b)t},\label{eq:diff}\end{equation}
and then note that 
 
\begin{equation}(1/t)\partial F(ut)/\partial u=(1/u)\partial F(ut)/\partial t
\equiv f(t|u)/u.\label{eq:clever}\end{equation}
Hence (\ref{eq:diff}) becomes
\begin{equation}g(t)=\frac{F(at)-F(bt)}{\ln(a/b)t}=\frac{\int_b^a
(f(t|u)/u)\dee u}{\int_b^a\dee u/u}.\label{eq:def}\end{equation}
The purpose of this paper is of course not to prove Frullani's
result, but note that (\ref{eq:frullani}) follows quickly from (\ref{eq:def})
on reversing the order of the two integrals.

Equation (\ref{eq:def}) shows that (\ref{eq:pdf}) is the pdf of a
mixture distribution generated by allowing the scale 
of a random variable from a survival distribution to be a random
variable $U$ in the range $(a,b)$, with pdf $\propto 1/u$. These
distributions have some attractive properties; for example, the
generation of random numbers is simple: one computes
$U=b(a/b)^X$, where $X \sim U[0,1]$, and then generates a random
number from the distribution with pdf $f$ with scale factor $U$.

Regarding notation, it is convenient to refer to $f$ as the
parent pdf, and $g$ as the daughter pdf, and to give daughter
distributions names such as the `$F_1$-gamma distribution'. The
parent pdfs are written as $f(t|u)$, where $u$ is the scale
factor, and the pdf with unit scale $f(t|1)$ is useful. The
scale parameters $(a, b)$ are omitted in daughter pdfs $g(t)$.
We write $r=a/b$ and for simplicity, standard distributions
where $b=1$ are given; it is easy to put the scale back in by $x
\rightarrow bx$. .  Finally, we use $T$ as the r.v. for
survival distributions, and $X$ for r.v.s defined on the whole
real line.

In general, the daughter distributions have longer tails than
their parent, and one extra parameter. These pdfs
are most attractive when the parent
distribution function can be written down explicitly, as is the case for
e.g.\ the exponential, Weibull and log-logistic distributions.
For the $F_1$-gamma  distribution, the pdf is expressible only
in terms of the incomplete gamma function.  When fitting censored
survival data, we also need the distribution function $G$
corresponding to the pdf $g$, or the survival function
$\bar{G}=1-G$, and this is often available as a special
function, and if not, can be evaluated as an integral.

The next section gives some properties of the $F_1$-distributions, then
$F_1$-distributions defined on the whole real line and multivariate
$F_1$-distributions are considered. Finally, some survival data with
covariates and two other datasets are fitted to exemplify the
use of the new distributions.
\section{Properties of the Mixture Distributions}
\subsection{Moments}
As with all mixture distributions, the moments are readily found
in terms of those of the parent. Let
$\text{E}(T^r)=\mu_r(u)$ for the parent distribution with 
scale factor $u$. Then for the pdf $g$
\[\text{E}(T^r)=\frac{\int_0^\infty \int_b^a f(t|u)t^r(\dee u/u)\dee t}{\ln(a/b)},\]
and reversing the order of integration,
\[\text{E}(T^r)=\frac{\int_b^a\mu_r(u)\dee u/u}{\ln(a/b)},\]
from which $\mu_r(a) < \text{E}(T^r) < \mu_r(b)$.
Since $\mu_r(u) \propto u^{-r}$, we have that
\begin{equation}\text{E}(T^r)=\frac{((a/b)^r-1)\mu_r(a)}{r\ln(a/b)}=\frac{(1-(b/a)^r)\mu_r(b)}{r\ln(a/b)}=\frac{(b^{-r}-a^{-r})\mu_r(1)}{r\ln(a/b)}.\label{eq:mom}\end{equation}

The coefficient of variation is related to $\text{CV}_p$, that
of the parent distribution, by
\[\text{CV}_d^2=\frac{(a+b)\ln(a/b)(CV_p^2+1)}{2(a-b)}-1,\]
from which $CV_d > CV_p$.

All moments of the daughter distribution are finite if those of
the parent are, and the moment generating function $M_g(s)$, if it exists, is 
\[M_g(s)=\ln(a/b)^{-1}\int_b^a M_f(s/u)\dee u/u,\]
where $M_f$ is the moment-generating function of the parent
distribution evaluated with unit scale.
For example, for an exponential parent distribution,
$M_f(s)=1/(1-s)$, so $M_g(s)=\ln(a/b)^{-1}\int_b^a
(1/(1-s/u))\dee u/u=\ln((a-s)/(b-s))/\ln(a/b)$.

\subsection{Relationship to slash distributions}
Tukey (\eg, Mosteller and Tukey, 1977) first introduced slash
distributions. The standard slash distribution is the distribution of the ratio
$T=X/V^{1/q}$, where $V$ is uniformly distributed in $(0,1)$,
and $X$ is for example normally distributed.
Taking now $V$ as uniform in $(b^q, a^q)$, we have that
\[G(t)=\text{Prob}(X/V^{1/q} < t)=\text{Prob}(X <
V^{1/q}t)=\frac{\int_{b^{1/q}}^{a^{1/q}}F(v^{1/q}t)\dee v}{a^q-b^q}.\]
Setting $u=v^{1/q}$ we have that
\[G(t)=q\frac{\int_b^aF(ut)u^{q-1}\dee u}{a^q-b^q}.\]
In the limit $q \rightarrow 0$
\[q/(a^q-b^q)=q/(\exp(q\ln(a))-\exp(q\ln(b))) \rightarrow 1/\ln(a/b)\]
and we have the distribution function corresponding to
(\ref{eq:def}). Hence the Frullani distributions are a type of
slash distribution.
\subsection{Other properties}
The survival function is
\[\bar{G}(t) = (\log a)^{-1} \int_t^{at} x^{-1} \bar{F}(x) \dee x\]
which leads via integration by parts to
\begin{equation}\bar{G}(t) = (\log r)^{-1}
       \{ \log(rt) \bar{F}(rt) - \log(t) \bar{F}(t)  
                         + \int_t^{rt} \log x \bar{F}(x) \dee x]. \label{eq:parts}\end{equation}
The tail behaviour of the $F_1$-distributions and some numerical
problems arising in inference are discussed in appendix A. It is
shown there that the hazard function behaves like that of the
parent with smallest scale factor, $b$. This can give hazard
functions that decrease in the tail, or bimodal hazard functions can occur.

In the left tail, as $t \rightarrow 0$, we have
\begin{equation}g(0)=\ln(a/b)^{-1}(a-b)f(0|1),\label{eq:zero}\end{equation}
where $f(0|1)$ is the parent pdf with unit scale factor. This
value of $g(0)$ is
intermediate between $f(0|b)$ and $f(0|a)$.

All daughter distributions examined from unimodal parent
distributions have been unimodal. It may be possible to produce
bimodal daughter distributions; a proof that daughter
distributions from unimodal parents must be unimodal is lacking.

Inference with covariates is straightforward.  To model
dependence on a vector of covariates ${\bf Z}$, one can use the
proportional hazards or additive hazards assumptions as usual,
\eg 
\[a=a_0\exp(-{\boldsymbol \gamma}^T{\boldsymbol Z}), \qquad b=b_0\exp(-{\boldsymbol \gamma}^T{\boldsymbol Z}),\]
where the overall scale is a function of the covariates, but the
range of scales $a/b$ is not.
\section{The Frullani-Weibull and Frullani-log-logistic distributions}
Brief details are given of the $F_1$-distributions for
two of the most important distributions used in survival analysis, the
Weibull and log-logistic.  The computation of the survival function is necessary
when fitting right-censored data by likelihood methods, so this
is also given. Moments are trivially derivable from the parent
distribution moments using (\ref{eq:mom}).

Substituting the Weibull
survival function $\bar{F}(t)=\exp(-t^\beta)$ into
(\ref{eq:spdf}) gives
\begin{equation}g(t)=\frac{\exp(-t^\beta)-\exp(-(rt)^\beta)}{\ln(r)t}.\label{eq:expweib}\end{equation}
The corresponding survival function $\bar{G}(t)$ is needed for
likelihood estimation with censored data. We have
\[\bar{G}(t)=\frac{\text{Ei}(-(rt)^\beta)-\text{Ei}(-t^\beta)}{\beta\ln(r)},\]
where $\text{Ei}(x)$ is the exponential integral
$\text{Ei}(x)=\int_{-\infty}^x\exp(y)\dee y/y$. Computation can
be done more easily using the related function $E_1$, where
$E_1(z)=\int_z^\infty \exp(-t)/t \dee t$. Then
\[\bar{G}(t)=\frac{\text{E}_1(t^\beta)-\text{E}_1((rt)^\beta)}{\beta\ln(r)}.\]
Although
$\bar{G}(t)$ cannot be expressed using elementary
functions, the exponential integral is a special function for
which there are well-established numerical approximations. The
$F_1$-Weibull with $\beta > 1$ has a hazard function that peaks and
then rises again in the tail.

Lifetime distributions can be mixed using the gamma density. The
exponential distribution then yields the Pareto distribution, so
the pdf (\ref{eq:expweib}) with $\beta=1$ is a shorter-tailed version of
the Pareto distribution. There is an application to frailty
analysis (\eg Klein and Moeschberger 2003); instead of using a
gamma density for the random hazard scaling factor $U$ within a
group, one can use the pdf $1/\ln(a/b)u$ for $b < U < a$.

For a log-logistic distribution with survival function
$S(t)=1/(1+t^\alpha)$, the pdf is
\[g(t)=\frac{(1+t^\alpha)^{-1}-(1+(rt)^\alpha)^{-1}}{\ln(r)t},\]
and on integrating
\[\bar{G}(t)=\frac{\ln\{(1+t^{-\alpha})/(1+(rt)^{-\alpha})\}}{\alpha\ln(r)}.\]
This distribution can have a bimodal hazard function, and is
used in section \ref{sec6} to exemplify the methodology.

The Frullani procedure can be repeated more than once, taking
(\ref{eq:pdf}) as the pdf for a second transformation. Here the
effect is to change the weighting of $f(t|u)$ from $\propto
1/u$. The resulting pdfs and distribution functions are complicated, and so this
second `Frullani-ization' has in general not been attempted.
However, for the log-logistic parent, the $n$th Frullani-ized
survival function can be written in terms of the $n$th
polylogarithm $\text{Li}_n$ (for polylogarithms, see \eg, Andrews
{\em et al}, 1999).

The survival function of the log-logistic distribution is
$\text{Li}_0(-t^{-\alpha})$, and from (\ref{eq:spdf}) the
survival function of the $F_n$ distribution is
obtained by
\[\text{Li}_{n-1}(-t^{-\alpha})\rightarrow
\frac{\text{Li}_n(-(rt)^{-\alpha})-\text{Li}_n(-t^{-\alpha})}{
\alpha\ln(1/r)}.\]
One can take varying scale ratios $r_n$ or
keep them all equal.
The pdf of the
$F_2$-log-logistic distribution is
\[g_2(t)=\frac{\ln\{\frac{(1+(r_1t)^{-\alpha})(1+(r_2t)^{-\alpha})}{(1+t^{-\alpha})(1+(r_1r_2t)^{-\alpha})}\}}{\alpha\ln(r_1)\ln(r_2)t}.\]
This distribution is very long-tailed.
The survival function is
\[\bar{G}_2(t)=\frac{\text{Li}_2(-t^{-\alpha})-\text{Li}_2(-(r_1t)^{-\alpha})-\text{Li}_2(-(r_2t)^{-\alpha})+\text{Li}_2(-(r_1r_2t)^{-\alpha})}{\alpha^2\ln(r_1)\ln(r_2)},\]
where the  dilogarithm function is
\begin{equation}\text{Li}_2(z)=-\int_0^z\ln(1-u)\dee u/u.\label{eq:di}\end{equation}

Further $F_1$-distributions are given in appendix B; some of these
are very tractable.
\section{Distributions defined on the whole real line}
One way to obtain distributions defined on the whole real
line is to transform the random variable to the positive real
line, \eg by taking its exponent, `Frullani-ize' the resulting
survival distribution, and then back-transform the result. In
terms of the original distribution function $F$, this yields
\[g(x)=(F(x+\alpha)-F(x-\alpha))/2\alpha,\]
where the mixing procedure consists simply of
making the centre of location a uniform random variable within
the range $(-\alpha,\alpha)$. Applying this to the normal
distribution yields a short-tailed distribution, with kurtosis
$\kappa=-\frac{6\alpha^4}{4(3+\alpha^2)^2}$. It becomes uniform
as $\alpha\rightarrow\infty$. This distribution was studied by
Bhattacharjee {\em et al} (1963). One can obtain other distributions using
different transformations from ${\cal R}^+$ to ${\cal R}$, such
as $X=T-1/T$, but the results are messy.

One can  also proceed by
reflecting (\ref{eq:def}) about the origin. This gives a smooth
distribution iff $\dee f(t|u)/\dee t=0$. For example, from the
half-normal distribution we obtain the $F_1$-Gaussian pdf
\begin{equation}g(x)=\frac{\Phi(a|x|)-\Phi(b|x|)}{\ln(a/b) |x|},\label{eq:symm}\end{equation}
where $\Phi$ is the standard normal distribution function. From
(\ref{eq:mom}), the variance is
\[\sigma^2=\frac{b^{-2}-a^{-2}}{2\ln(a/b)},\]
and the kurtosis is
\[\kappa=3\frac{(a/b)^2+1}{(a/b)^2-1}\ln(a/b)-3,\]
an increasing function of $a/b$. The value of $a/b$
corresponding to a kurtosis $\kappa$ can be found by Newton-Raphson
iteration. It is best to start at a value $\ge 3$, because the
slope $\dee \kappa/\dee (a/b)=0$ at $a/b=1$, and low slopes can
cause the method to diverge.

The pdf at $x=0$ from (\ref{eq:zero}) is $\frac{(a-b)}{\ln(a/b)\sqrt{2\pi}}$.

The moment-generating function is
\begin{equation}M_g(s)=(1/2)\ln(a/b)^{-1}\{\text{Ei}(s^2/2b^2)-\text{Ei}(s^2/2a^2)\}.\label{eq:em}\end{equation}
The distribution
function is probably best evaluated as
\[G(x)=1/2+\text{sign}(x) \ln(a/b)^{-1}\int_{b|x|}^{a|x|}\{\Phi(x)/x\} \dee x.\]

\begin{figure}[b]
\centering
\makebox{\includegraphics{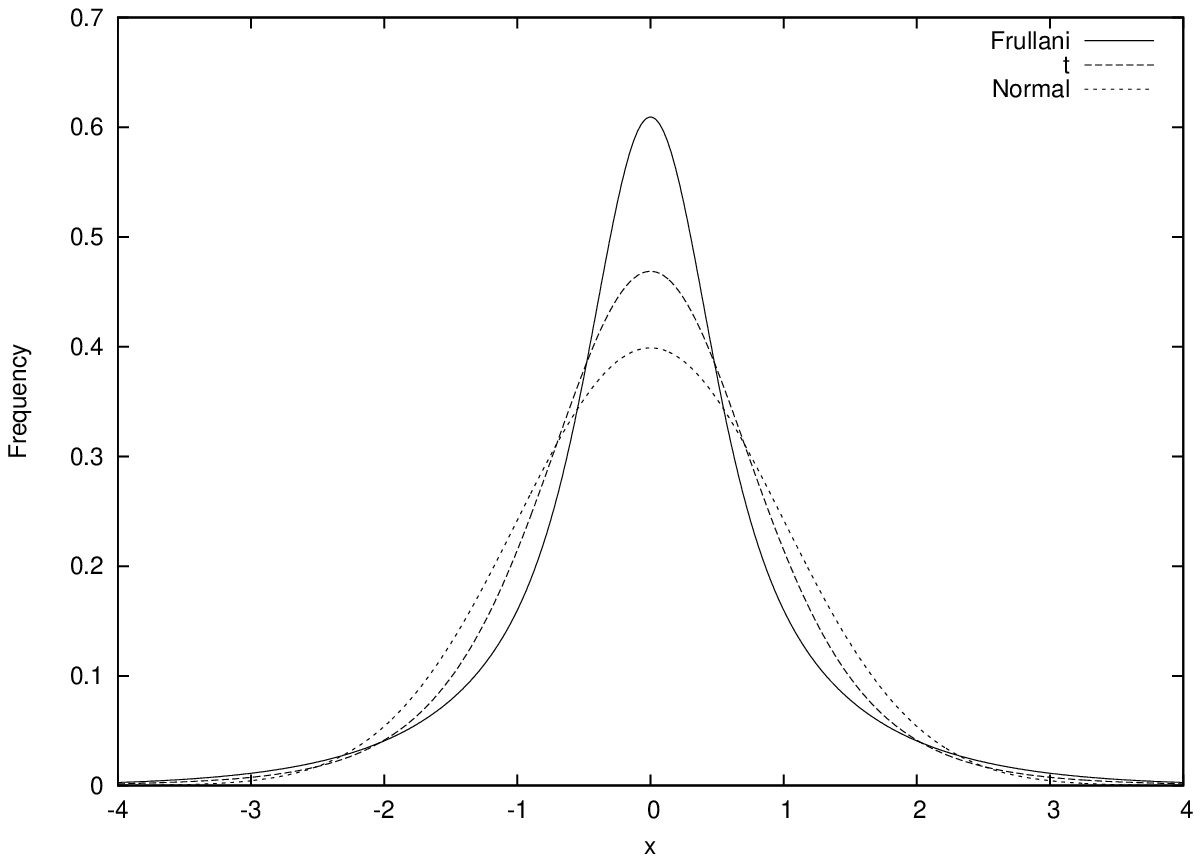}}
\caption{\label{figa} Frullani-normal, t  probability density functions with zero
mean, unit variance, and kurtosis 3, and Gaussian pdf.}
\end{figure}

\begin{figure}
\centering
\makebox{\includegraphics{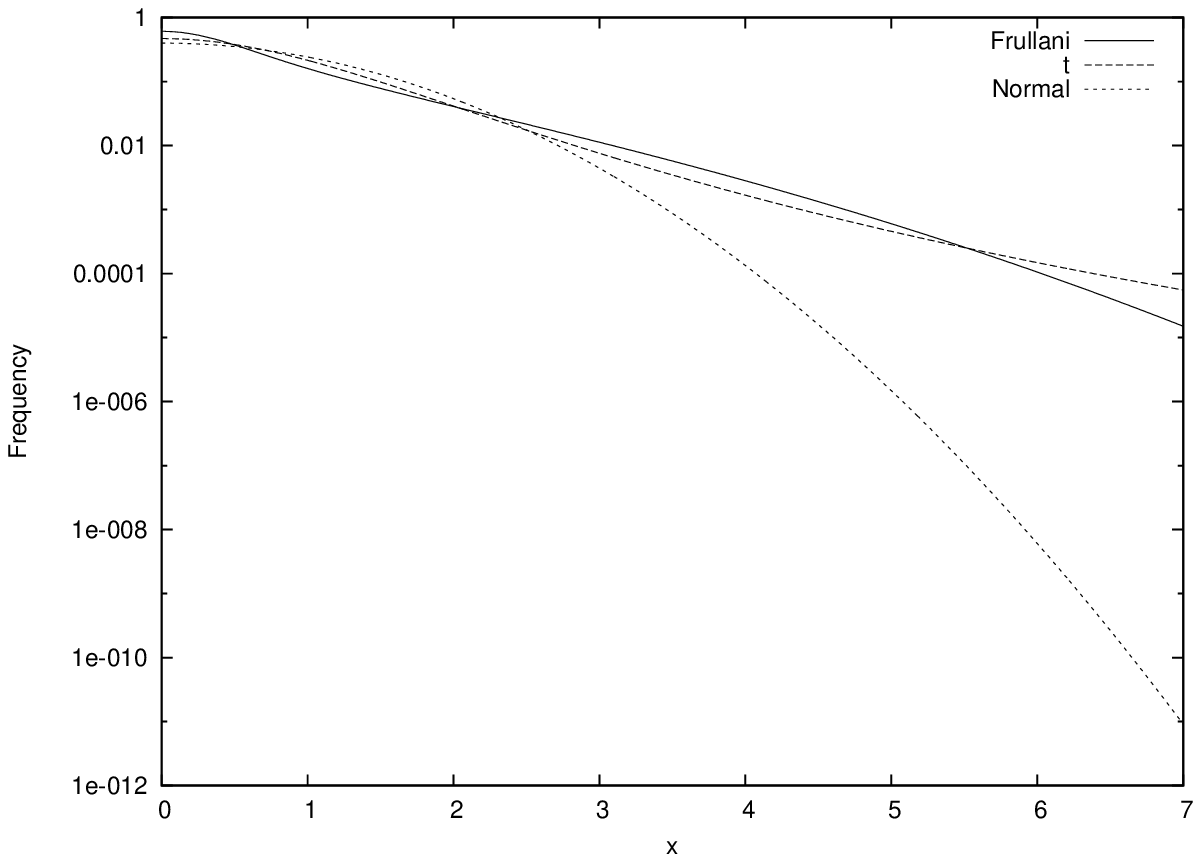}}
\caption{\label{figb}Tail behaviour of Frullani-normal, t probability density functions with zero
mean, unit variance, and kurtosis 3, with Gaussian pdf.}
\end{figure}

This distribution is long-tailed, with the normal as a special case
when $b=a$. Figure \ref{figa} shows this distribution and the
t-distribution, both with unit variance and kurtosis $\kappa=3$.
It can be seen that this distribution is heavier in the tail
than the t-distribution. Figure \ref{figb} shows the tail
behaviour, where the t-distribution is heavier in the extreme tail.

Skewness can be introduced by a device used by Azzalini (\eg Azzalini
and Capitanio, 1999, but used earlier by {\em inter alia} O'Hagan and Leonard
(1976)), when for example
\begin{equation}g(x)=2\Phi(\lambda x)\frac{\Phi(a|x|)-\Phi(b|x|)}{\ln(a/b) |x|},\label{eq:phis}\end{equation}
where $\lambda$ can be of either sign.

Writing the function of $|x|$ as a mixture, changing the
order of integration, applying the method of parts and making
the substitution $u=\lambda\tan(\theta)$ yields the mean
\[\mu=\frac{2\{(1+(\lambda/b)^2)^{1/2}-(1+(\lambda/a)^2)^{1/2}\}}{\sqrt{2\pi}\ln(a/b)\lambda}.\]
The second moment $\text{E}(X^2)$ is that of (\ref{eq:symm}), so
that the variance is $\frac{b^{-2}-a^{-2}}{2\ln(a/b)}-\mu^2$,
and the skewness can then be found from the third moment $\text{E}(X^3)$,
obtained like the mean as
\[\frac{2\{\frac{2}{3}(1+(\lambda/b)^2)^{3/2}-\frac{2}{3}(1+(\lambda/a)^2)^{3/2}+(\lambda/a)^2(1+(\lambda/a)^2)^{-1/2}-(\lambda/b)^2(1+(\lambda/b)^2)^{-1/2}\}}{\sqrt{2\pi}\ln(a/b)\lambda^3}.\]
The kurtosis can be found from the fourth moment
\[\text{E}(X^4)=\frac{\frac{3}{4}(b^{-4}-a^{-4})}{\sqrt{2\pi}\ln(a/b)},\] 
that of (\ref{eq:symm}).

There is however a more natural way of skewing the $F_1$-normal
distribution, without going beyond the concept of `Frullani-ization'.
One can generate half-normal pdfs with parameters $(a,b)$ for $X
> 0$ and $(a,c)$ for $X < 0$. The two pdfs are made to be equal
at $X=0$ by weighting them, so that
\begin{equation}g(x)=\left\{ \begin{array}{ll}
2s\frac{\Phi(ax)-\Phi(bx)}{\ln(a/b)x} & \text{if $x > 0$} \\
2(1-s)\frac{\Phi(a|x|)-\Phi(c|x|)}{\ln(a/c)|x|} & \text{if $x < 0$} 
\end{array}
\right., \label{eq:two}\end{equation}
where 
\[s=\frac{(a-c)/\ln(a/c)}{(a-b)/\ln(a/b)+(a-c)/\ln(a/c)}.\]
This is the probability that $X$ exceeds the mode.
The pdf (\ref{eq:two}) is that of a two-piece distribution.
 As the matching occurs at
the mode $x=0$ the first derivative of $g$ is also
continuous.

The moments are simple if messy functions of $a, b, c$ and
$\mu$, if the distribution is translated to have its mode at $\tilde{\mu}$.
We have then
\[\text{E}((X-\tilde{\mu})^{2n})=\frac{(2n-1)!!}{2n}\{\frac{s(b^{-2n}-a^{-2n})}{\ln(a/b)}+\frac{(1-s)(c^{-2n}-a^{-2n})}{\ln(a/c)}\},\]
and
\[\text{E}((X-\tilde{\mu})^{2n+1})=\frac{2^{n+1}n!}{\sqrt{2\pi} (2n+1)}\{\frac{s(b^{-2n-1}-a^{-2n-1})}{\ln(a/b)}-\frac{(1-s)(c^{-2n-1}-a^{-2n-1})}{\ln(a/c)}\}.\]
In particular,
\[\text{E}(X)-\tilde{\mu}=\frac{\sqrt{2}(a-b)(a-c)(c-b)}{\sqrt{\pi}abc\{(a-b)\ln(a/c)+(a-c)\ln(a/b)\}}.\]
This distribution looks promising in its ability to fit skew and
long-tailed data. The moments are finite and easily calculable,
and a nice feature for data fitting and inference is that the mode is a
parameter of the distribution. 

If the distribution is skew to
the right, $c < b$ and the probability $s$ of exceeding the mode
$> 1/2$. There is an inferential problem for the skew-t
distribution in that the log-likelihood is bimodal as a function
of $\lambda$; there is no such difficulty with the two-piece
distribution proposed here.

Two-piece distributions, often based on the normal distribution,
appear in the literature occasionally, for example the `two-piece
normal' distribution of Gibbons and Mylroie (see \eg Johnson,
Kotz and Balakrishnan (1994), p173). Here the standard
deviations differ in each half. Jones (2006) also discusses
general two-piece distributions where the scale differs in the
two halves. Note that (\ref{eq:two}) is not of this type,
although one could construct such an $F_1$-distribution from (\ref{eq:pdf}).
\section{Multivariate distributions}
The simple expression (\ref{eq:def}) for $g(t)$ extends simply
to the multivariate case. To illustrate with the bivariate case,
let the mixture distributions for two time measures $X$ and $Y$
be defined over $(a,b), (c,d)$ respectively. Then we define the
bivariate pdf
\[g(x,y)=\frac{\int_b^a\int_d^c \frac{\partial^2F(ux,vy)}{\partial
x\partial y}(\dee u/u)(\dee v/v)}{\ln(a/b)\ln(c/d)},\]
where $F$ is a bivariate distribution function.
Applying (\ref{eq:clever}), this becomes
\[g(x,y)=\frac{\int_b^a\int_d^c(\partial^2F(ux,vy)/\partial u\partial
v)\dee u\dee v}{\ln(a/b)\ln(c/d)xy},\]
so that finally
\[g(x,y)=\frac{F(ax,cy)-F(bx,cy)-F(ax,dy)+F(bx,dy)}{\ln(a/b)\ln(c/d)xy}.\]
This by the way gives a bivariate result analogous to the
Frullani integral:
\begin{multline}\int_0^\infty\int_0^\infty \frac{f(ax,cy)-f(bx,cy)-f(ax,dy)+f(bx,dy)}{xy}=\\
\ln(a/b)\ln(c/d)\{f(\infty,\infty)-f(0,\infty)-f(\infty,0)+f(0,0)\}.\end{multline}

The mixing does not induce a correlation between
$X$ and $Y$, but a bivariate survival distribution where $X$ and
$Y$ are correlated can be modified in this way to give longer
tails, the degree of tail elongation being allowed to differ
between the two variables. This property is often required. In
terms of copulae, the dependence parameter of the copula is unchanged.

The flexibility to give differing tail lengths is achieved with
two extra parameters. Clearly, multivariate extensions are
straightforward.

Elliptical distributions (\eg Johnson and Kotz, 1972) are
becomingly increasingly popular, and another possible way to
extend the methodology to the multivariate case is to take a
mixture of elliptical distributions.  For example, the
$p$-variable  normal pdf, with a scaling factor of $u$ is
\[f({\boldsymbol x}|u)=\frac{u^p\exp(-u^2{\boldsymbol
x}^T{\boldsymbol  V}^{-1}{\boldsymbol
x}/2)}{(2\pi)^{p/2}|{\boldsymbol  V}|^{1/2}}.\]
Allowing $U$ to have a gamma distribution yields the
multivariate t-distribution (Johnson and Kotz, 1972). The
corresponding $F_1$-multivariate normal distribution has pdf
\[g({\boldsymbol x})=\frac{\int_b^a u^{p-1}\exp(-u^2{\boldsymbol
x}^T{\boldsymbol  V}^{-1}{\boldsymbol
x}/2)\dee u}{(2\pi)^{p/2}\ln(a/b)|{\boldsymbol  V}|^{1/2}}.\] One
parameter is redundant, so we can set \eg $a=1$. Hence
\[g({\boldsymbol x})=\frac{\Gamma(p/2)\{F(Q,p/2)-F(b^2Q,p/2)\}}{\pi^{p/2} \ln(1/b)Q^{p/2}|{\boldsymbol  V}|^{1/2}},\]
where $Q={\boldsymbol x}^T{\boldsymbol  V}^{-1}{\boldsymbol
x}$ and $F(Q,p/2)$ is the distribution function of the
chi-squared distribution with $p$ degrees of freedom.
For example, when $p=2$, the bivariate pdf is
\[g({\boldsymbol x})=\frac{\exp(-b^2Q/2)-\exp(-Q/2)}{\pi \ln(1/b)|{\boldsymbol  V}|^{1/2}Q}.\]
The moment-generating function $\text{E}\exp({\boldsymbol
s}^T{\boldsymbol X})$ is
\[M_g({\boldsymbol s})=(1/2)\ln(a/b)^{-1}\{\text{Ei}({\boldsymbol
s}^T{\boldsymbol  V}{\boldsymbol s}/2b^2)-\text{Ei}({\boldsymbol  s}^T{\boldsymbol
V}{\boldsymbol  s}/2)\}\exp({\boldsymbol  \xi}^T{\boldsymbol  s}),\]
where ${\boldsymbol \xi}$ is the mean, analogously to (\ref{eq:em}).

In general, the moments of the mixed distributions are given by
\[\text{E}(X^rY^s)=\frac{b^{-(r+s)}-1}{r+s}\mu_{rs},\]
where $\mu_{rs}$ is the corresponding moment of the parent distribution.

These distributions can of course be skewed \eg by Azzalini's
method (Azzalini and Capitanio, 1999, 2003), as is done for the multivariate t-distribution.
Some more $F_1$-distributions are described in appendix B.
\section{\label{sec6}Results of Data Fitting}
To illustrate the use of Frullani-mixed distributions in survival
analysis, a dataset from Klein and Moeschberger 2003, section 1.14 was analysed.
The times in weeks to weaning of first-born babies from 927 young mothers
were found from interview, along with some covariates, such as
race (white, black, or other), whether the mother smoked at the
birth of the child, years of mother's schooling, whether the
mother lived in poverty, and so on. Censoring is light; some
mothers stopped breastfeeding before the baby was weaned (\ie,
switched to bottle-feeding, etc). The
dataset is available from the authors' website.

\begin{figure}
\centering
\makebox{\includegraphics{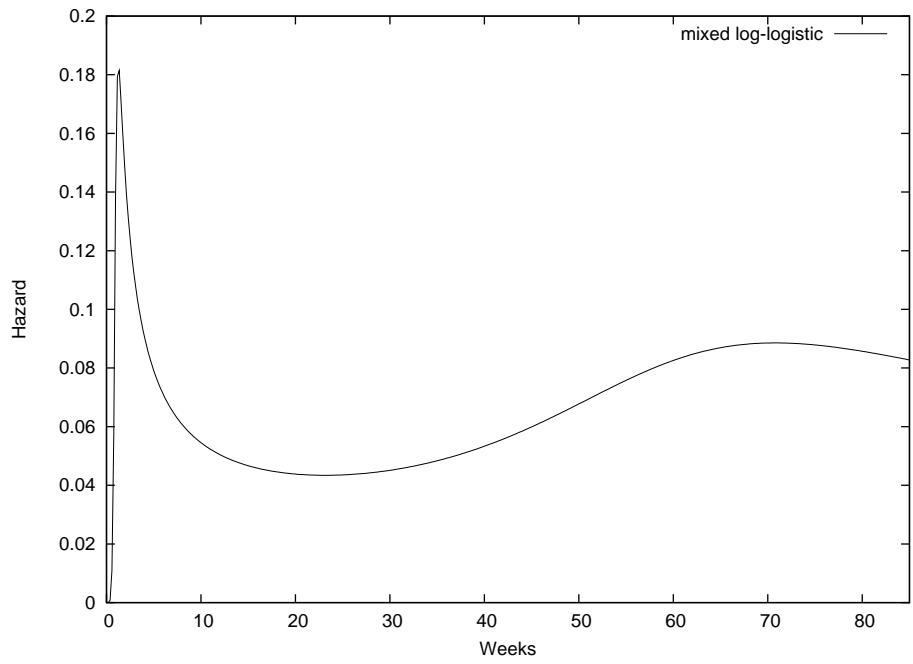}}
\caption{\label{figc} Frullani-mixed log-logistic hazard
function for the weaning dataset from Klein and Moeschberger.}
\end{figure}

There is considerable evidence of rounding of the number of
weeks to weaning, but the data are usable. Table \ref{tab:a}
shows minus log-likelihoods ($-\ell$) for fits of the models
discussed above to the data, with no covariates. It can be seen
that the mixture models fit considerably better than
conventional models. The scale ratio $a/b$ is high, showing a
considerable departure from the parent distribution.
Figure \ref{figc} shows the hazard function, of the
Frullani-mixed log-logistic model ($F_1$-log-logistic model) ,
with 95\% confidence intervals.

The hazard function decreases and then rises again after about 50
weeks. This was also noted by Klein and Moeschberger, and is the
reason why the mixture models perform so well here. After a tail
of slow weaning, most mothers still breastfeeding wean the baby
after a year, only a very few continuing longer. This type of
behaviour, with a long tail which however eventually peters out,
is well fitted by these mixture models.

Table \ref{tab:b} shows covariate parameter estimates, standard
errors (or coefficient of variation) and 95\% confidence
intervals, for the mixed log-logistic model. The log-likelihood
increased by 7.84 on adding the covariates, some of which are
statistically significant. The findings obtained here are
similar to those of Klein and Moeschberger.

\begin{figure}
\centering
\makebox{\includegraphics{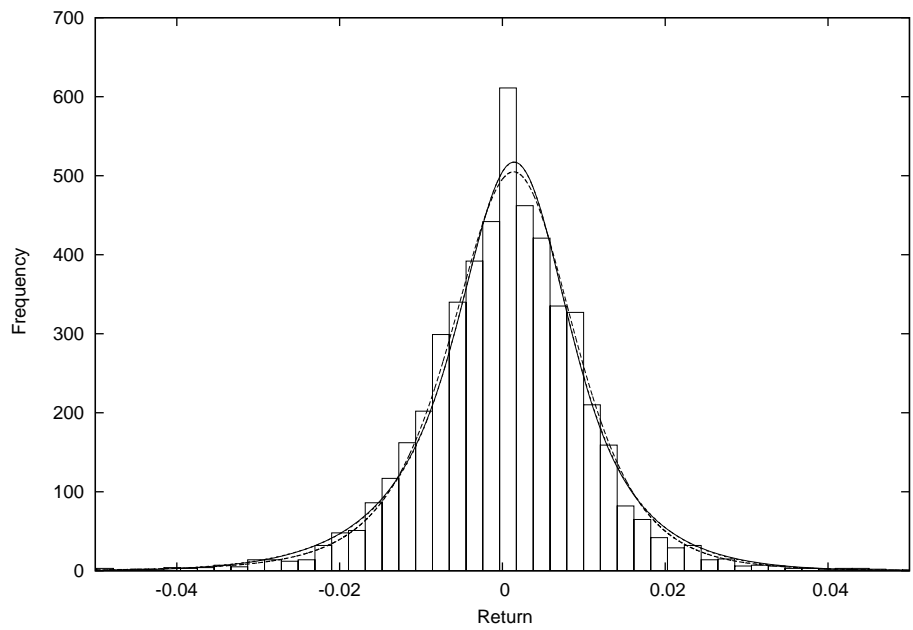}}
\caption{\label{figd} Skew Frullani-normal distribution fitted
to FTSE 100 daily (continuously compounded) returns from 1984 to
2003. The dot-dashed line is the Azzalini skew-t distribution. The
two-piece $F_1$-normal distribution cannot be distinguished from the
skew-$F_1$-normal distribution.}
\end{figure}

To illustrate the use of (\ref{eq:phis}) and (\ref{eq:two}),
5106 FTSE-100 returns from 1984 to 2003 were fitted by maximum
likelihood estimation.  Figure \ref{figd} shows the model fit,
and that of the Azzalini skew-t distribution. This latter has
4.3 degrees of freedom, corresponding to a long tail. However,
(\ref{eq:phis}) and (\ref{eq:two}) fitted the data with a nearly
identical log-likelihood. The fitted ratio a/b was 4.41 for
(\ref{eq:phis}). As the skewness is small here, the lines from
these two pdfs can not be distinguished.

\begin{figure}
\centering
\makebox{\includegraphics{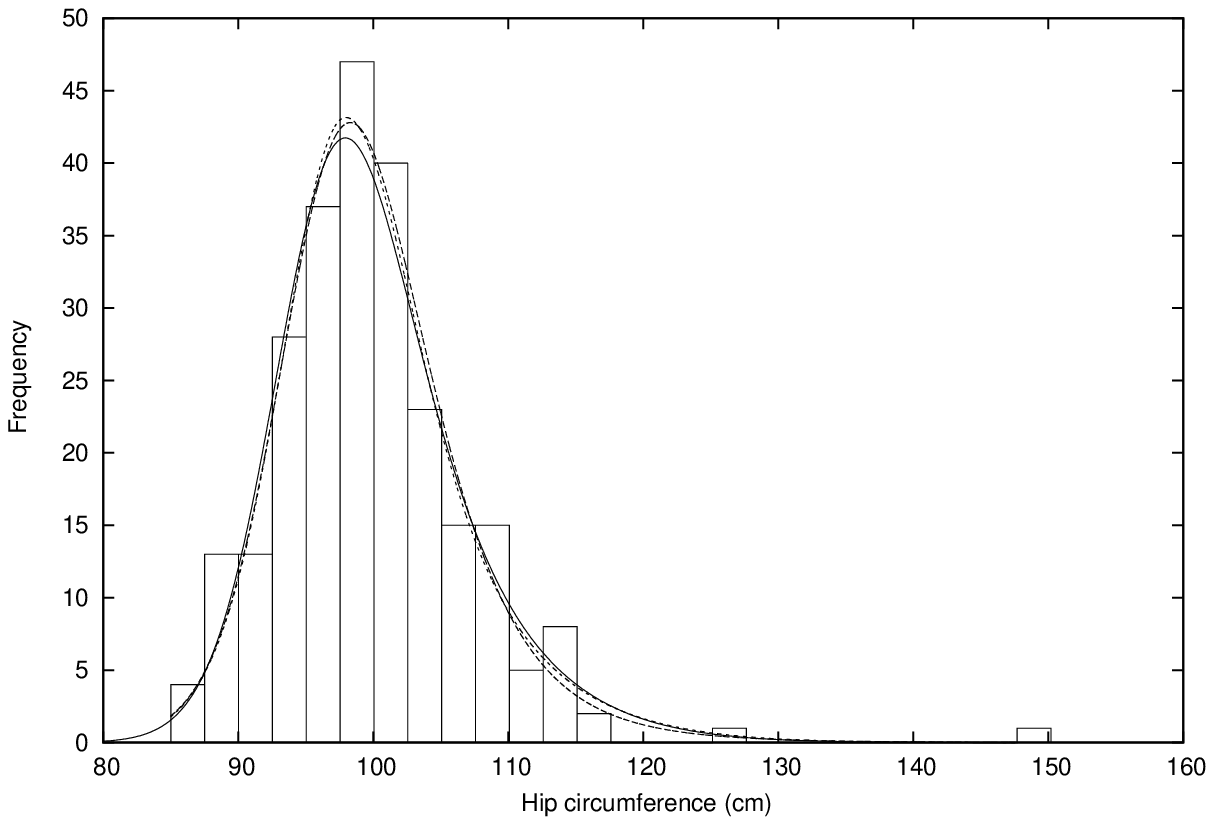}}
\caption{\label{fige} Skew Frullani-normal distribution fitted
to hip size measurements. The dot-dashed line is the Azzalini skew-t
distribution, and the dotted line the two-piece $F_1$-normal distribution.}
\end{figure}

Finally, figure \ref{fige} shows the fit of this distribution to
some hip circumference data from the Statlib database. The dataset contains estimates
of the percentage of body fat determined by underwater weighing
and various body circumference measurements for 252 men, and was
submitted to Statlib by Roger Johnson. 

Here the data are also long-tailed, the skew-t distribution
fitting with 5.8 degrees of freedom.
The pdf (\ref{eq:phis}) fits with an identical
log-likelihood function, and a ratio $a/b$ of 4.84, and the
two-piece $F_1$-normal distribution fitted very similarly.

\begin{figure}
\centering
\makebox{\includegraphics{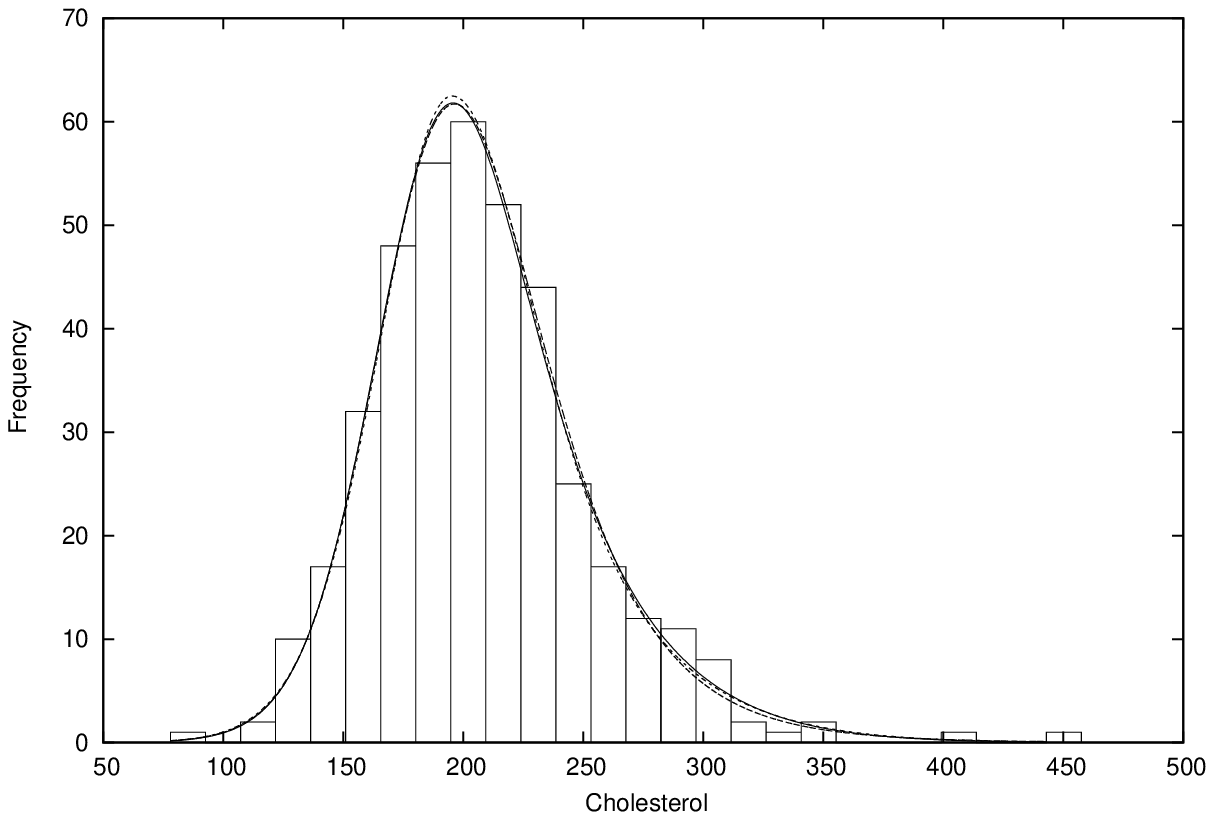}}
\caption{\label{figf} Skew Frullani-normal distribution fitted
to cholesterol measurements. The dot-dashed line is the Azzalini skew-t
distribution, and the dotted line the two-piece $F_1$-normal distribution.}
\end{figure}

Figure \ref{figf} shows
fits to cholesterol level of 403 patients interviewed in
connection with diabetes, given in Harrell (2001). The Azzalini
skew distribution fitetd with $\nu=6.7$ degrees of freedom, and
all three fits had very similar log-likelihood and overlap on
the plot.

\section{Conclusions}
The Frullani integral (\ref{eq:frullani}) has a probabilistic
interpretation, \ie that, given a pdf, a mixture distribution where the
logarithm of the scale of the random variable is uniformly
distributed within a finite range, is also a pdf. The Frullani
integral leads to a class of `daughter' distributions, which are
here called `$F_1$-distributions', whose pdfs are simply expressible
in terms of the `parent' distribution functions.  

That this is so is theoretically interesting; the mathematical result has
been known since 1821, but never applied to
`distribution-ology'. However, this result will doubtless be of little
interest to statisticians, unless the new distributions obtained
are practically useful.  They are relatively tractable: the moments
are simply expressible in terms of the parent distribution moments, and
random numbers are readily generated if they can be for the
parent distribution. 

The most direct application is to survival distributions. Here there
is already a wide range of univariate distributions available,
with hazard functions capturing most of the behaviour actually
observed. Hazards can rise or fall (Weibull model), rise then
fall (lognormal and log-logistic models), or fall then rise, the
`bathtub' hazard of human mortality and some equipment failure.
There are some particular new distributions that might be useful,
such as the $F_1$-exponential and $F_1$-Weibull (\ref{eq:expweib}). But
what is unusual about the new class of distributions is the tail
behaviour. In the tail, the hazard becomes that of the parent
distribution with the smallest scale factor (largest mean). For
a parent distribution such as the lognormal, the hazard function
can be bimodal.  This type of behaviour has been observed, \eg
in breast cancer (Demicheli {\em et al} 2008), in bovine foetal
survival (Hanson {\em et al} 2003), and in the weaning example
used in this paper. In general, these distributions can be very
long-tailed, but behave like the parent distribution in the
extreme tail. For example, if $b \ll a$, in the tail
$\bar{F}(at) \ll \bar{F}(bt)$, so from (\ref{eq:spdf}) $g(t) \sim
1/\ln(a/b)t$, until in the extreme tail $g(t) \sim \bar{F}(bt)/\ln(a/b)t$.

In finance, this type of behaviour has to be
engineered, via the `truncated Levy' distribution. Finite
moments are desirable, and to achieve this Ali and Nadarajah
(2006) truncated the Pareto distribution, and Nadarajah (2009) has
produced truncated versions of five distributions used in
finance, telecommunications \etc, including the t-distribution,
with the aim of achieving finite moments.  Hopefully, many more
examples of this type of behaviour will come to light; in many
instances, distributions can be long tailed or heavy tailed,
but often there is eventually some kind of limitation.  For
example, human weight is ultimately limited by physics and
biology.


By exploiting the reflection symmetry of distributions such as
the Gaussian, an $F_1$-Gaussian distribution can be derived. Such
distributions, and their multivariate generalisation, provide an
alternative to the t-distribution for which all moments exist.
The two-piece skewed distribution is one of a useful class of
distributions that can be skew and long-tailed, with the normal
distribution as a special case.

\section*{Acknowledgement}The author is grateful to Prof. Chris
Jones for pointing out the connection with slash distributions
and for helpful comments on the manuscript.

\section*{Appendix A: Some more properties of the $F_1$-distributions}
\subsection*{Tail behaviour}
The hazard function of the pdf $g(t)$ is
\[h_g(t)=\frac{\int_b^a h(t|u)\bar{F}(ut)\dee u/u}{\int_b^a \bar{F}(ut)\dee u/u}.\]
Writing 
\[\bar{F}(ut)=\exp(-H(ut))=\exp(-\int_0^t h(v|u)\dee v),\]
we have that
\[h_g(t)=\frac{\int_b^a h(t|u)\exp(-(H(ut)-H(bt)))\dee u/u}{\int_b^a \exp(-(H(ut)-H(bt)))\dee u/u}.\]
If $H(ut)-H(bt)=\int_{bt}^{ut} h(v|1) \dee v \rightarrow\infty$
as $t \rightarrow\infty$, then $h_g(t) \rightarrow h(t|b)$,
because terms with $u > b$ have weight zero. Probabilistically,
in the tail, events from distributions with $u > b$ have already
occurred, and only the longest-tailed distribution in the
mixture can still supply events.

However, if
$H(ut)-H(bt)$ tends to a constant, all components of the mixture
contribute with constant weight. The hazard function $h(t|u)$,
derived from $\bar{F}(ut)=\exp(-H(bt)-c)$ is then independent of
$u$, so we can again write $h_g(t) \rightarrow h(t|b)$.

Finally, $H(ut)-H(bt)$ cannot tend to zero, as we would not
then have $\bar{F}(ut) \rightarrow 0$ as $t \rightarrow\infty$.
Hence it is always true that $h_g(t) \rightarrow h(t|b)$.  This
means that a distribution such as the Weibull with an increasing
hazard function gives rise to a mixture distribution where the
hazard first increases, then decreases, and finally increases
again in the extreme tail.

An example where $H(ut)-H(bt)$ tends to a constant is the Pareto
distribution $\bar{F}(ut)=1/(1+ut)$, so long-tailed that the
mean is not defined. Here $H(ut)=\log(1+ut) \rightarrow
\log(ut)$, and $H(ut)-H(bt)=\ln(u/b)$. The hazard function
$h(t|u) \rightarrow 1/t$.

\subsection*{Numerical problems}
These mixture distributions have scale parameters $a$ and
$b$, as well as any others intrinsic to the parent distribution,
such as the Weibull shape parameter.  When maximising likelihood
functions, the function minimizer may choose values such that $b
> a$. Although the pdf is invariant under $a \leftrightarrow b$,
the log-likelihood function may then contain logarithms of
negative argument. One can resolve this problem by using parameters
$\alpha_1, \alpha_2$, and taking
$a=\text{max}(\alpha_1,\alpha_2)$,
$b=\text{min}(\alpha_1,\alpha_2)$, or by taking parameters $a$
and $r=b/a < 1$.
When $\ln(a/b) < \epsilon$, computation of the pdf and survival
function become numerically unstable, and one simply computes the
parent pdf and survival function. For the two-piece skew
distribution (\ref{eq:two}) there are three scale parameters,
$a, b, c$. It is probably best to take $a, r_1=b/a$ and
$r_2=c/a$ as parameters, where $0 < r_1 < 1$, $0 < r_2 < 1$. One
has to test for $|ax-bx| < \epsilon$ and $|ax-cx| < \epsilon$, and revert to
the parent pdf if so.

A deeper problem is
now identified. Consider (\ref{eq:pdf}) when $a=b+\delta$ and
$\delta \ll c$. Expanding $F$ in a McLaurin series in $\delta$,
and writing $F_n\equiv\dee^n F((b+\delta)x)/\dee \delta^n|_{\delta=0}$,
\[g(x) \simeq \frac{F_1\delta+(1/2)F_2\delta^2+(1/6)F_3\delta^3}{x\{\delta/b-(1/2)(\delta/b)^2+(1/3)(\delta/b)^3\}},\]
The log-likelihood of observing an uncensored sample $x_1 \ldots
x_n$ is
\[\ell(b)\simeq\sum_{i=1}^n\{\ln(f(x_i|b)+\ln(1+(f_3/2F_1)\delta+(F_3/6F_1))-\ln(1-(1/2)(\delta/b)+(1/3)(\delta/b)^2)\},\]
where the first term is $f=(b/x)F_1$.
Hence
\[\ell(b) \simeq \sum_{i=1}^n\{ \ln(f(x_i|b)+(1/2)(F_2/F_1+1/b)\delta+(F_3/6F_1-F_2^2/8F_1^2-5/24b^2)\delta^2\}.\]
The linear term in $\delta$ 
vanishes, by virtue of the fact that $\hat{b}$ is a MLE;
$\sum_{i=1}^n (F_2/F_1+1/b)=\sum_{i=1}^n \partial
f(x_i|b)/\partial b=0$.
This shows that there is a small computational problem for inference, if we
 maximise the log-likelihood for the parent distribution,
and restart a function minimizer at the point $b=\hat{b},
\delta=0$. At that point $\partial \ell/\partial\delta=0$, so
some minimizers could `stick' and exit with $\hat{\delta}=0$.
\subsection*{Score tests}
Consider (\ref{eq:pdf}) when $a=c+\delta$, $b=c-\delta$, and
$\delta \ll c$.  The central differencing of the distribution
function $F$ allows us to explore `sensible' hypotheses, where
we have a maximum-likelihood fit to  data of the parent pdf
$f$, with scale parameter $c$, and we consider taking a mixture
of scale factors around $c$. Expanding in a McLaurin series in $\delta$,
and writing $F_n=\dee^n F((c+\delta)x)/\dee \delta^n|_{\delta=0}$,
\[g(x) \simeq \frac{F_1\delta+(1/6)F_3\delta^3+(1/120)F_5\delta^5}{x\{(\delta/c)+(1/3)(\delta/c)^3+(1/5)(\delta/c)^5\}}.\]
The log-likelihood on observing an uncensored sample 
is
\[\ell(c)\simeq\sum_{i=1}^n\{\ln(f(x_i|c)+\ln(1+(f_3/6F_1)\delta^2+(F_5/120F_1))-\ln(1+(1/3)(\delta/c)^2+(1/5)(\delta/c)^4)\},\]
where the first term is $f=(c/x)\partial F/\partial c$.
Hence
\[\ell(c) \simeq \sum_{i=1}^n \{\ln(f(x_i|c)+((F_3/6F_1)-(1/3c^2))\delta^2+((F_5/120F_1)-(F_3^2/72F_1^2)-13/90c^4)\delta^4\}.\]
From this the score statistic
$\partial\ell/\partial\delta|_{\delta=0}$ can be read off as
\[\partial\ell/\partial\delta|_{\delta=0}=(1/3)\sum_{i=1}^n\{F_3/2F_1-1/c^2\}\]
or
\[\partial\ell/\partial(\delta^2)|_{\delta=0}=(1/6)\hat{c}^2\sum_{i=1}^n
x_i^2 f''(x_i)/f(x_i)-2n,\]
where $f''$ denotes the second derivative w.r.t. $x_i$.
The variance of the score is 
\[\text{E}\{-\partial^2\ell/\partial(\delta^2)^2|_{\delta=0}\}=n\text{E}\{F_3^2/36F_1^2-F_5/60F_1+13/45c^4\},\]
which can be approximated as the sample sum. Asymptotically the
standardised score is $\sim N[0,1]$ under $H_0$.

For the exponential distribution, the score statistic is
$(1/6)\{\sum_{i=1}^n(x_i-\bar{x})^2-\bar{x}^2\}$. The score test
is a one-sided test of whether the standard deviation exceeds
the mean, this equality being a well-known property of the
exponential distribution.

\section*{Appendix B: More distributions}
\subsection*{The $F_1$-gamma distribution}
The parent pdf is
$f(t|1)=t^{\beta-1}\exp(-t)/\Gamma(\beta)$, with
distribution function 
\[F(t)=\Gamma(\beta)^{-1}\int_0^{t}x^{\beta-1}\exp(-x)\dee x=P(\beta,t)=1-Q(\beta,t),\]
where $P$ is the incomplete gamma function.
The daughter distribution is then
\[g(t)=\frac{P(\beta,rt)-P(\beta,t)}{\ln(r)t},\]
with survival function
\[\bar{G}(t)=(\ln(r))^{-1}\int_{t}^{rt}Q(\beta,x)\dee x/x.\]
This can be integrated by parts using (\ref{eq:parts}) to give the single integral
\[\bar{G}(t)=(\ln(r))^{-1}\{Q(\beta,rt)\ln(rt)-Q(\beta,t)\ln(t)\}+\Gamma(\beta)^{-1}\int_{t}^{rt}x^{\beta-1}\exp(-x)\ln(x)\dee
x.\]
The mean for the general distribution is
\[\text{E}(T)=\mu=(b^{-1}-a^{-1})\beta/\ln(a/b),\]
variance
\[\text{var}(T)=\beta(\beta+1)(b^{-2}-a^{-2})/2\ln(a/b)-\mu^2.\]
\subsection*{The $F_1$-lognormal distribution}
With $\ln(X)$ standard normal, the
mixture pdf is
\[g(t)=\frac{\Phi(\ln(rt))-\Phi(\ln(t))}{\ln(r)t},\]
where $\Phi$ is the standard normal distribution function.

On using (\ref{eq:parts}) we obtain the survival function
\[\bar{G}(t)=1+\frac{\frac{\{\exp(-\ln^2(t)/2)-\exp(-\ln^2(rt)/2)\}}{\sqrt{2\pi}}-(\ln(rt)\Phi(\ln(rt))-\ln(t)\Phi(\ln(t)))}{\ln(r)}.\]
This fortunately requires only the computation of elementary
functions and the error function.
\subsection*{The $F_1$-Pareto distribution}
Taking the parent survival function as
$\bar{F}(t)=(1+t)^{-\alpha}$ for $T > 0$, the $F_1$-Pareto pdf
is
\[g(t)=\frac{(1+t)^{-\alpha}-(1+rt)^{-\alpha}}{\ln(r)t}.\]

The survival function is
\[\bar{G}(t)= (\alpha\ln(r))^{-1}\{t^{-\alpha}
{_2F_1}(\alpha,\alpha;\alpha+1;-1/t))-(rt)^{-\alpha} {_2F_1}(\alpha,\alpha;\alpha+1;-1/rt))\}.\]

\subsection*{The $F_1$-Cauchy distribution}
The distribution function corresponding to the Cauchy pdf
$f(x|1)=\frac{1}{\pi(1+x^2)}$ is
$F(t)=1/2+\frac{\tan^{-1}x}{\pi}$, from which the $F_1$-Cauchy pdf
derived from Frullani-izing the right half of the distribution
and reflecting about the origin is
\[g(x)=\frac{\tan^{-1}(rx)-\tan^{-1}(x)}{\pi\ln(r)x},\]
or
\[g(x)=\frac{\tan^{-1}\{(r-1)x/(1+rx^2)\}}{\ln(r)\pi x}.\]
The distribution function is
\[G(x)=1/2+i\text{sign}(x)\{2\pi\ln(r)\}^{-1}\{\text{Li}_2(-irx)-\text{Li}_2(irx)-\text{Li}_2(-ix)+\text{Li}_2(ix)\},\]
where $\text{Li}_2$ is the dilogarithm defined by (\ref{eq:di}).
\section*{Figures and Tables}
\clearpage
\begin{table}
\caption{Log-likelihoods and fitted parameter values for
survival models fitted to the weaning times
from Klein and Moeschberger (no covariates included).
\label{tab:a}}
\begin{tabular}{|l|c|c|c|c|} \hline
Model & -Log-likelihood& Shape param. & $a$ & $b$ \\ \hline
Exponential & 3409.29 & 1 & 0.0594826 & n/a \\
Mixed exponential & 3406.36 & 1 & 0.108615 & 0.0359083 \\  
Weibull  & 3408.56 & 0.970074 & 0.0602816 & n/a \\  
Mixed Weibull & 3388.41 & 2.01358 & 0.482310 & 0.0152277 \\  
Gamma & 3409.27 & 0.992376 & 0.0590219 & n/a \\  
Mixed gamma & 3379.93 & 5.35736 & 3.90543 & 0.0865375 \\  
Lognormal & 3402.77 & 1.17603 & 0.106397 & n/a \\  
Mixed lognormal & 3374.38 & 0.403807 & 0.856643 & 0.0173032 \\  
Log-logistic & 3429.32 & 1.43847 & 0.101974 & n/a \\  
Mixed log-logistic & 3372.66 & 7.38682 & 1.06599 & 0.0159815 \\  \hline
\end{tabular}
\end{table}

\begin{table}
\caption{Parameter estimates for the analysis of weaning data
from Klein and Moeschberger, with standard errors (or coefficient
of variation where marked), and 95\% confidence interval. \label{tab:b}}
\begin{tabular}{|l|c|c|c|} \hline
Parameter & estimate & SE & 95\% CI \\ \hline
Shape $\alpha$ & 6.60933 & 0.2308 (CV) & (4.204491   10.389665) \\  
Scale $a$ & 1.70841 & 0.2101 (CV) & (1.131659    2.579107) \\  
Scale $b$ & 0.027619 & 0.2512 (CV) & (0.016881    0.045187) \\  
Race: Black & -.155922 & 0.07564 & (-0.304182   -0.007663) \\ 
Race: Other & -.0009 & 0.09047 & (-0.178226    0.176426) \\  
Smoked at birth & -.109407 & 0.06312 & (-0.233113    0.014300) \\  
Years schooling & 0.0440546 & 0.01878 & (0.007238    0.080871) \\  
Poverty & 0.196038 & 0.07658 & (0.045936    0.346140) \\  \hline
\end{tabular}
\end{table}

\end{document}